\documentstyle[stwol,epsf]{article}

\def\plb#1{Phys.~Lett.~{\bf B#1}}
\def\npb#1{Nucl.~Phys.~{\bf B#1}}
\def\prl#1{Phys.~Rev.~Lett.~{\bf #1}}
\def\prd#1{Phys.~Rev.~{\bf D#1}}
\def\zpc#1{Z.~Phys.~{\bf C#1}}

\def\l{\left}
\def\r{\right}

\def\la{\langle}
\def\ra{\rangle}

\def\im{\mbox{Im}\,}

\def\tab#1{Table~\ref{#1}}

\def\fig#1{Fig.~\ref{#1}}

\def\eq#1{Eq.~(\ref{#1})}

\def\re#1{Ref.~\cite{#1}}

\def\br#1{{\cal B}\l(#1\r)}

\newdimen\unit
\def\point#1 #2 #3{\vbox to0pt{\kern-#2\unit
  \hbox{\kern#1\unit#3}\vss}
 \nointerlineskip}

\newcommand{\be}{\begin{equation}}
\newcommand{\ee}{\end{equation}}
\newcommand{\bea}{\begin{eqnarray}}
\newcommand{\eea}{\end{eqnarray}}

\newcommand{\gev}{\mbox{ GeV}}

\begin{document}

\begin{titlepage}

\begin{center}

\renewcommand{\thefootnote}{\fnsymbol{footnote}}

{\Large\bf Centre de Physique Th\'eorique~\footnote{
Unit\'e Propre de Recherche 7061
}, CNRS Luminy, Case 907\\
F-13288 Marseille -- Cedex 9}

\vspace{3 cm}

{\huge\bf Lattice-Constrained Dispersive Bounds 
for $\bar B^0\to\pi^+\ell^-\bar\nu$ Decays~\footnote{Talk given at 
the 28$^{th}$ International Conference on High Energy Physics 
(ICHEP96), 
Warsaw, Poland, 25-31 July 1996. To appear in the proceedings.}}

\vspace{0.7 cm}

{\Large\bf 
Laurent Lellouch}~\footnote{{\it email: lellouch@cpt.univ-mrs.fr}}

\vspace{2,3 cm}

{\bf Abstract}

\end{center}

I present a recent piece of work on semileptonic $B\to\pi$ decays in
which lattice results and kinematical and dispersive constraints are
combined to obtain model-independent bounds on the relevant form
factors and rates.

%\vspace{2,3 cm}
\vfill

\noindent Key-Words: Semileptonic Decays of $B$ Mesons, Determination of 
Kobayashi-Maskawa Matrix Elements ($V_{ub}$), Dispersion Relations,
Lattice QCD Calculation, 
Heavy Quark Effective Theory.

\bigskip

\setcounter{footnote}{0}
\renewcommand{\thefootnote}{\arabic{footnote}}

\noindent Number of figures: 1

\bigskip

\noindent September 1996

\noindent CPT-96/P.3385
\bigskip

\noindent anonymous ftp or gopher: cpt.univ-mrs.fr
\end{titlepage}

\title{Lattice-Constrained Dispersive Bounds 
for $\bar B^0\to\pi^+\ell^-\bar\nu$ Decays}

\author{Laurent Lellouch}

\address{Centre de Physique Th\'eorique (UPR 7061), CNRS-Luminy, 
Case 907, F-13288 Marseille, France}


\twocolumn[\maketitle\abstracts{
I present a recent piece of work on semileptonic $B\to\pi$ decays in
which lattice results and kinematical and dispersive constraints are
combined to obtain model-independent bounds on the relevant form
factors and rates.}]

\section{Introduction}

The CLEO Collaboration has very recently presented a measurement of the
branching ratios for $B\to\pi\ell\bar\nu$
decays ($\ell{=}e,\mu$)~\cite{cleo}:
\be
\br{\bar B^0\to\pi^+\ell^-\bar\nu}=(1.8\pm0.4\pm0.3\pm0.2)\times 10^{-4}
\ ,
\label{eq:brcleo}
\ee
where the errors are statistical, systematic and the estimated
model-dependence introduced in determining efficiencies.
This measurement represents an excellent opportunity to determine the
poorly know CKM matrix element $|V_{ub}|$. This
determination requires calculating, over the full kinematical
range, the non-perturbative matrix element:
\bea
\la \pi^+(p')|V^\mu|\bar B^0(p)\ra &= 
\l(p+p'-q\frac{M^2-m^2}{q^2}\r)^\mu
f^+\nonumber\\
&+ q^\mu\frac{M^2-m^2}{q^2}f^0
\ ,\eea
where $q{=}p{-}p'$, $V^\mu{=}\bar u\gamma^\mu b$, M is the mass of the
$B$ and $m$, that of the $\pi$.  
Though this matrix element can be
calculated using lattice QCD, current day lattice
simulations, with lattice spacings on the order of $3\gev^{-1}$, 
cannot cover the full
kinematical range. The problem is that the energies and momenta of the
particles involved, whose orders of magnitude are set by the $b$ quark
mass ($m_b\simeq 5\gev$), are large on the scale of the cutoff over much
of phase space.  To
limit these energies in relativsitc lattice quark calculations, one
performs the simulation with heavy-quark mass values $m_Q$ around that
of the charm ($m_c\simeq 1.5\gev$), where discretization errors remain
under control. Then one extrapolates the results up to $m_b$ by
fitting heavy-quark scaling relations (HQSR) with power corrections to
the lattice results.  Another approach is to work with discretized
versions of effective theories such as Non-Relativistic QCD (NRQCD) or
Heavy-Quark Effective Theory (HQET) in which the mass of the heavy
quark is factored out of the dynamics. All these approaches, however,
are based on the heavy-quark expansion which is more limited
for $heavy\to light$ quark decays, such as the one that concerns us
here, than it is for $heavy\to heavy$ quark decays: it imposes no
normalization condition on the relevant form factors at the the zero
recoil point $q^2=q^2_{max}$ and only applies in a limited region
around $q^2_{max}$. Furthermore, momentum-dependent discretization
errors restrict the momentum of the initial and final state mesons to
around $1\sim 2\gev$. Thus, all these approaches are constrained to
relatively small momentum transfers: one can only reconstruct the
$q^2$ dependence of the relevant form factors in a limited region
around $q^2_{max}$ and one is left with the problem of extrapolating
these results to smaller $q^2$.

$heavy\to light$ quark decays are difficult in any theoretical
approach. Indeed, they require understanding the underlying QCD dynamics
over a large range of momentum transfers from $q^2_{max}{=}26.4\gev^2$
for semileptonic $B\to\pi$ decays, where
the $\pi$ is at rest in the frame of the $B$ meson, to
$q^2{=}0$ where it recoils very strongly.

\section{$\bar B^0\to\pi^+\ell^-\bar\nu$ and Dispersive Constraints}

The solution to the problem of the limited kinematical reach of
lattice simulations of $heavy\to light$ quark decays presented here
consists in supplementing lattice results for the relevant form
factors around $q^2_{max}$ with dispersive bound techniques to obtain
improved, model-independent bounds for the form factors for all
$q^2$.~\cite{btopi}  For the case of $\bar B^0\to\pi^+\ell^-\bar\nu$
decays, one can use the kinematical constraint, $f^+(0)${=}$f^0(0)$,
to further constrain the bounds.

\subsection{Dispersive Bounds}

The subject of dispersive bounds in semileptonic decays 
has a long history going back to S. Okubo {\it et al.}~who
applied them to semileptonic $K\to\pi$ decays~\cite{SOkS71}. 
C. Bourrely {\it et al.}~first combined these techniques with
QCD and applied them to semileptonic $D\to K$ decays \cite{CBoMR81}. 
Very recently,
C.G. Boyd {\it et al.}~applied them to $B\to\pi\ell\bar\nu$
decays \cite{CBoGL95}. 

The starting point for $B\to\pi\ell\bar\nu$ decays is the polarization
function
\bea
\Pi^{\mu\nu}(q){=}i\int d^4x\ e^{iq\cdot x} \la 0|T\l(V^\mu(x)
V^{\nu\dagger}(0)\r)|0\ra\nonumber\\
{=}(q^\mu q^\nu-g^{\mu\nu}q^2)\,\Pi_T(q^2)+q^\mu q^\nu \,\Pi_L(q^2)
\ ,\label{twopoint}
\eea
where $\Pi_{T(L)}$ corresponds to the propagation of a
$J^P{=}1^-\,(0^+)$ particle. The corresponding spectral functions,
$\im\Pi_{T,L}$, are sums of positive contributions coming from
intermediate $B^*$ ($J^P{=}1^-$), $B\pi$ ($J^P{=}0^+\mbox{ and }1^-$),
$\ldots$ states and are thus upper
bounds on the $B\pi$ contributions. 
Combining, for instance, the bound from $\im\Pi_L$ with the
dispersion relation ($Q^2{=}-q^2$)
\bea
\chi_L(Q^2)&=&\frac{\partial}{\partial Q^2} (Q^2\Pi_L(Q^2))\nonumber\\ 
&=&\frac{1}{\pi}\int_0^\infty
dt\frac{t\,\im\Pi_L(t)}{\l(t+Q^2\r)^2}
\ ,\label{eq:disprelS}
\eea
one finds
\bea
\chi_L(Q^2)\ge\frac{1}{\pi}\int_{t_+}^\infty
dt\,k(t,Q^2)|f^0(t)|^2
\ ,\label{eq:chilbnd}
\eea
where $t_{\pm}{=}(m_B\pm m_\pi)^2$ and $k(t,Q^2)$ is a kinematical
factor.  Now, since $\chi_L(Q^2)$ can be calculated analytically in
QCD for $Q^2$ far enough below the resonance region (i.e. $-Q^2\ll
m_b^2$), \eq{eq:chilbnd} gives an upper bound on the weighted integral
of the magnitude squared of the form factor $f^0$ along the $B\pi$
cut. To translate this bound into a bound on $f^0$ in the region of
physical $B\to\pi\ell\bar\nu$ decays is a problem in complex
analysis (please see \re{btopi} for details).
A similar constraint can be obtained from $\Pi_T$ for $f^+$. There, however,
one has to confront the additional difficulty that $f^+$ is not analytic
below the $B\pi$ threshold because of the $B^*$ pole.

The beauty of the methods of \re{CBoMR81} is that they enable one to
incorporate information about the form factors, such as their
values at various kinematical points, to constrain the bounds.
For the case at hand, however, these methods must be generalized in two
non-trivial ways. In constructing these generalizations, one must
keep in mind that the bounds: 1) form inseparable pairs; 2) do not 
indicate the probability that the form factor
will take on any particular value within them.

\subsection{Imposing the Kinematical Constraint}

The first problem is that \eq{eq:chilbnd} and the equivalent
constraint for $f^+$ yield independent bounds on the form
factors which do not satisfy the kinematical constraint 
$f^+(0){=}f^0(0)$. The bounds on $f^+$ require $f^+(0)$ to lie within
an interval of values $I_+$ and those on $f^0$, within an interval
$I_0$.  Together with these bounds, however, the kinematical constraint 
requires $f^+(0){=}f^0(0)$ to lie somewhere within $I_+\cap I_0$.
Thus, we seek bounds on the form factors which are consistent
with this new constraint.

A natural definition is to require these new bounds to be the envelope
of the set of pairs of bounds obtained by allowing $f^+(0)$ and
$f^0(0)$ to take all possible values within the interval $I_+\cap
I_0$.  In \re{btopi}, it is shown how this envelope can be constructed
efficiently and that the additional constraint can only improve the
bounds on the form factors for all $q^2$. Also, as a by product, one
obtains a formalism which enables one to constrain bounds on a
form factor with the knowledge that it must lie within an interval of
values at one or more values of $q^2$.

\subsection{Taking Errors into Account}

As they stand, the methods of \re{CBoMR81} can only accommodate exact
values of the form factors at given kinematical points and contain no
provisions for taking errors on these values into account. Of course,
the results given by the lattice do carry error bars. More precisely,
the lattice provides a probability distribution for the value of the
form factors at various kinematical points. What must be done, then,
is to translate this distribution into some sort of probability
statement on the bounds.
The conservative solution is to consider the probability that complete
pairs of bounds lie within a given finite interval at each value of
$q^2$. Then, using this new probability, one can define upper and
lower $p\%$ bounds at each $q^2$ as the upper and lower boundaries of
the interval that contains the central $p\%$ of this 
probability.\footnote{The density of pairs of bounds increases 
toward the center 
of the distribution as long as the distribution of the lattice results
does.} These bounds indicate that there is at least a $p\%$ probability that
the form factors lie within them at each $q^2$.

\subsection{Lattice-Constrained Bounds}

\begin{figure}[tb]
\setlength{\epsfxsize}{64mm}\epsfbox[80 250 470 545]{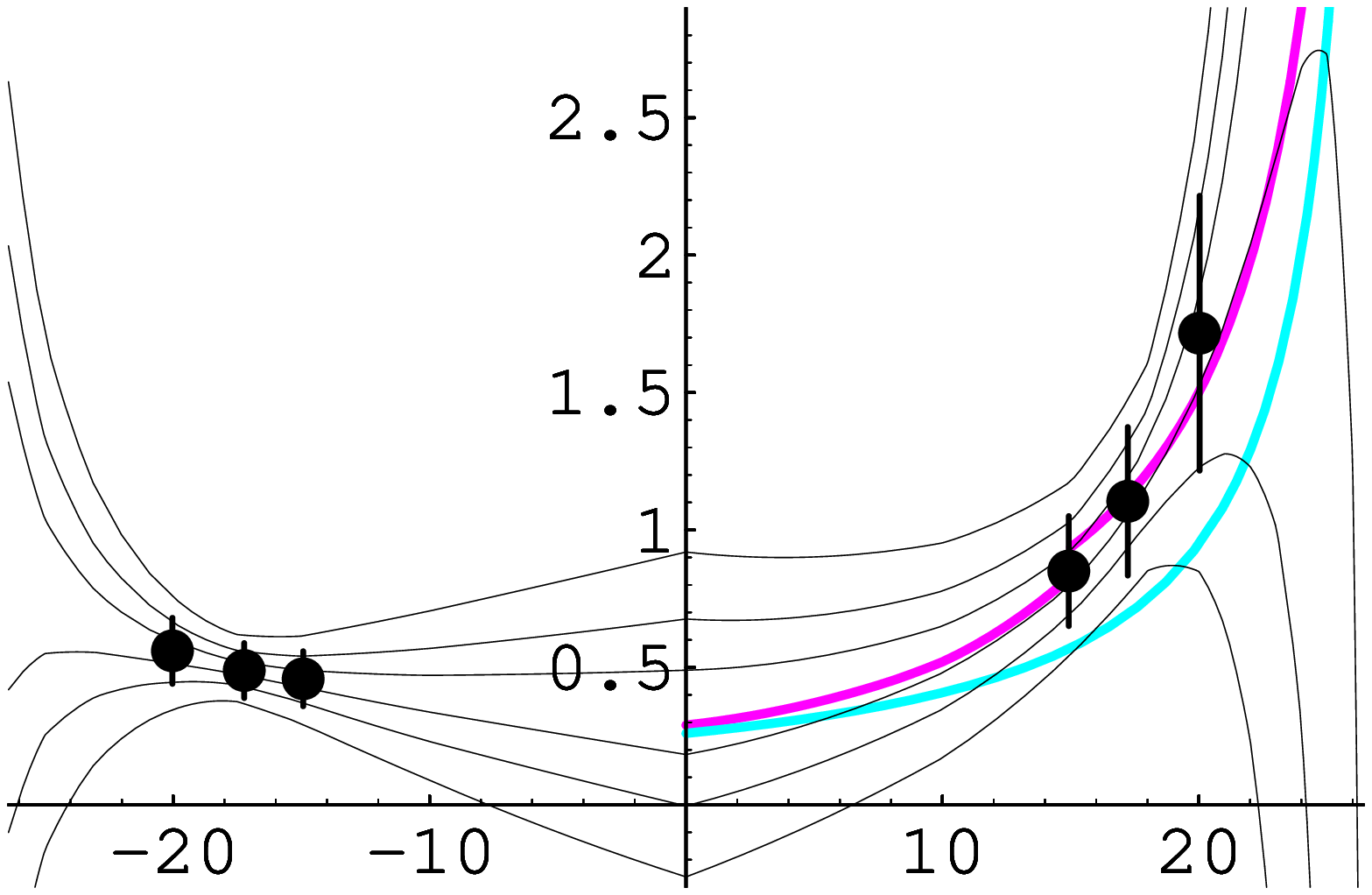}
\unit=0.8\hsize
\point 0.95 0.05 {\large{$q^2(\mbox{GeV}^2)$}}
\point 0.20 0.75 {\Large{$f^0(|q^2|)$}}
\point 0.770 0.75 {\Large{$f^+(q^2)$}}
\caption{$f^0(|q^2|)$ and $f^+(q^2)$ versus $q^2$.
The data points are the lattice results of UKQCD~\protect\cite{DBuetal95} 
with added systematic errors. 
The pairs of fine curves are, from the outermost
to the innermost, the 95\%, 70\% and 30\% bounds. The upper shaded curve is 
the LCSR result of \protect\re{VBeBKR95} and the lower one, the SR3 results of 
\protect\re{PaB93}.}
\label{fig:lcsr}
\end{figure}

To constrain the bounds on $f^+$ and $f^0$, the lattice
results of the UKQCD Collaboration
\cite{DBuetal95} are used, to which a large range of systematic
errors is added to ensure that the bounds obtained are conservative.
Because of these systematic errors, the probability distribution of
the lattice results is not known. The simplifying and rather
conservative assumption that the results are uncorrelated and gaussian
distributed is made.  The required probability is constructed by
generating 4000 pairs of bounds from a Monte-Carlo on the distribution
of the lattice results.  The results for the bounds on the form factors
are shown in
\fig{fig:lcsr}. The two form factors are plotted back-to-back to
show the effect of the kinematical constraint. Without this constraint, 
the bounds
on $f^+$ would be looser, especially around $q^2{=}0$, where phase
space is large. Since $f^+$ determines the rate, the kinematical 
constraint and the bounds on $f^0$ are important.

Also shown in \fig{fig:lcsr} is the light-cone sumrule (LCSR) result
of \re{VBeBKR95} and the 3-point sumrule (SR3) result of \re{PaB93}.
The latter is parametrized by a pole form with $f^+(0)=0.26$ and
$m_{pole}=5.25\gev$~\cite{PaB93}. The former has two components: for
$q^2$ below $15\gev^2$, the $q^2$ dependence of $f^+$ is determined
directly from the sumrule; for larger $q^2$, pole dominance is assumed
with a residue determined from the same correlator. While agreement of
the LCSR result with the bounds is excellent, the SR3 result is quite
strongly disfavored. The bounds are compared with the
predictions of more authors as well as with direct fits of various
parametrizations to the lattice results in \re{btopi}.  Though again certain
predictions are strongly disfavored, the lattice results and bounds
will have to improve before a firm conclusion can be drawn as to the
precise $q^2$ dependence of the form factors.

The bounds on $f^+$ also enable one to constrain the $B^*B\pi$
coupling $g_{B^*B\pi}$ which determines the residue of the $B^*$ pole
contribution to $f^+$.  The constraints obtained are poor because
$f^+$ is weakly bounded at large $q^2$, as can be seen in
\fig{fig:lcsr}. Fitting the lattice results for $f^0$ and $f^+$ to a
parametrization which assumes $B^*$ pole dominance for $f^+$ and which
is consistent with HQS and the kinematical constraint gives the more
precise result $g_{B^{*+}B^o\pi^+}=28\pm 4$.\footnote{The result of
this fit is entirely compatible with our bounds on $f^+$ and $f^0$.}
However, because this result is model-dependent, it should be taken
with care.

\subsection{Bounds on the rate and $|V_{ub}|$}

As was done for the form factors, one can define the probability of
finding a complete pair of bounds on the rate in a given interval and
from that probability determine confidence level (CL) intervals for the
rate.  The resulting bounds are summarized in \tab{tab:rate}. They
were obtained by appropriately integrating the 
4000 bounds generated for $f^+(q^2)$, 
taking the skewness of the resulting ``distribution'' 
of bounds on the rate into account. Also given in \tab{tab:rate}
are bounds on $f^+(0)$ as well as the preditions of other authors
for these quantities. Although the bounds on $f^+(q^2)$ disfavor
some of these predictions rather strongly, the bounds on the rate
only exclude the quark model result of \re{NaIS89} with high
certainty. 
\begin{table}[tb]
\caption{Bounds on rate in units of $|V_{ub}|^2\,ps^{-1}$ and on $f^+(0)$:
top block. Quark model (QM) and light-cone (LCSR), 2pt (SR2) and 3pt
(SR3) sumrule predictions: bottom block.
\label{tab:rate}}
\begin{tabular}{ccc}
\hline
$\Gamma\l(\bar B^0\to\pi^+\ell^-\bar\nu\r)$ & $f^+(0)$ & details\\
\hline
$2.4\to 28$ & $-0.26\to 0.92$ & 95\% CL \\
$2.8\to 24$ & $-0.18\to 0.85$ & 90\% CL \\
$3.6\to 17$ & $0.00\to 0.68$ & 70\% CL \\
$4.4\to 13$ & $0.10\to 0.57$ & 50\% CL \\
$4.8\to 10$ & $0.18\to 0.49$ & 30\% CL \\
\hline
$7.4\pm 1.6$ & $0.33\pm 0.06$ & QM~\cite[(WSB)]{WSB85} 
\\ 
2.1 & 0.09 & QM~\cite[(ISGW)]{NaIS89}\\
9.6 & & QM~\cite[(ISGW2)]{DaSI95}\\
& $0.26$ & QM~\cite{PaOX95}\\
& $0.24-0.29$ & QM~\cite{ChCHZ96}\\
9.6--15.2 & $0.29-0.46$ & QM~\cite{IGNS96}\\
$7\pm2$ & $0.20-0.29$ & QM~\cite{DMe96}\\
$14.5\pm 5.9$ & $0.4\pm 0.1$ & SR2~\cite{CeDP88}\\
4.5--9.0 & $0.27\pm 0.05$ & SR$2+3$~\cite{AO89}\\
$3.60\pm 0.65$ & $0.23\pm 0.02$ & 
SR3~\cite{StN95} \\
$5.1\pm 1.1$ & $0.26\pm 0.02$ & 
SR3~\cite{PaB93} \\
$8.1$  & $0.24-0.29$ & LCSR~\cite{VBeBKR95}\\
$7\pm 1$ & 0.21 -- 0.27 & LAT~\cite[(UKQCD)]{DBuetal95}\\
$9\pm 6$ & 0.10--0.49 & LAT~\cite[(ELC)]{ELC94}\\
$8\pm 4$ & 0.23--0.43 & LAT~\cite[(APE)]{apebpi}\\
\hline
\end{tabular}
\end{table}

The CL bounds of \tab{tab:rate} can also be used, in conjunction with
the branching ratio measurement of CLEO given in \eq{eq:brcleo}, to
determine $|V_{ub}|$. One finds
\be
|V_{ub}|10^4\sqrt{\tau_{B^0}/1.56\,ps}=(34\div 49)\pm 8\pm 6
\ ,\label{eq:vub}
\ee
where the range given in parentheses is that obtained from the
30\% CL bounds on the rate and represents the most probable
range of values for $|V_{ub}|$. The first set of errors is obtained
from the 70\% CL bounds and the second is obtained by
combining all experimental uncertainties in quadrature and applying them to
the average value of $|V_{ub}|$ given by the 30\% CL results.
This determination of $|V_{ub}|$ has a theoretical error of approximately
37\%. Though non-negligible, this error is quite reasonable given
that the bounds on the rate are completely model-independent
and are obtained from lattice data which lie in a limited kinematical domain 
and include a conservative range of systematic errors. For comparison, 
the lattice results of ELC~\cite{ELC94} and APE~\cite{apebpi}
were obtained from the determination of $f^+$ at a single 
$q^2\sim 18-20\gev^2$
supplemented with the assumption of $B^*$-pole dominance while those
of UKQCD~\cite{DBuetal95} rely on constrained model fits to the same
values of $f^+$ and $f^0$ (without the added
systematics errors) used to obtain the bounds presented here.

\section{Conclusion and Outlook}

I have presented a new formalism by which lattice results for
semileptonic $B\to\pi$ decays, limited to a narrow kinematical range, are
combined with dispersive and kinematical constraints to obtain
model-independent bounds on the relevant form factors and rates over
the full kinematical domain. I have compared these bounds, which 
have a well defined statistical meaning, to the predictions of other authors.

Though the bounds 
will benefit from forthcoming, improved
lattice results, they would benefit most from an increase in the range 
of $q^2$. 

Finally, the techniques presented here
are in principle applicable to limited results obtained by non-lattice means
as well as to other processes such as $B\to\rho\ell\bar\nu$
and $B\to K^*\gamma$ decays.

\end{document}